\documentclass{article}

\usepackage[sorting=none, style=numeric-comp]{biblatex} 

\usepackage{graphicx} 

\usepackage{tikz}

\usepackage{amsmath}
\usepackage{amssymb}
\usepackage{stmaryrd}
\usepackage{hyperref}
\usepackage{fullpage}

\usepackage[dvipsnames]{xcolor}

\newcommand{\tikzcircle}[2][red,fill=red]{\tikz[baseline=-0.5ex]\draw[#1,radius=#2] (0,0) circle ;}%

\newcommand{\openmist}{\tikzcircle[black, fill=white]{5pt}}

\newcommand{\filledmist}{\tikzcircle[black, fill=black]{5pt}}

\newcommand{\ket}[1]{\lvert #1\rangle}

\newcommand{\braket}[2]{\langle #1\mid #2\rangle}

\newcommand{\den}[1]{\llbracket #1\rrbracket}

\newcommand{\question}[1]{\textbf{#1}}

\newcommand{\Norm}{\mathcal{N}}

\title{Certified Misty-State Rewriting\\[0.25em]
\large A Question-and-Answer Guide}

\author{
  \textbf{Tony Newton} \\
  Newton Astro Labs\thanks{ Newton Astro Labs is the trading name under which Tony Newton conducts independent computational research in the United Kingdom; it is not a limited company.} \\
  London, UK \\
  \texttt{tony.newton79@gmail.com}
  \and
  \textbf{Dan-Adrian German} \\
  Indiana University \\
  Bloomington, IN, USA \\
  \texttt{dgerman@iu.edu}
}

\date{}

\begin{document}

\maketitle

\begin{abstract}
Quantum mechanics is difficult to teach because its conceptual content and mathematical notation usually arrive together. Rudolph's misty-state language was designed
to decouple those burdens; basis states are visual objects, clouds represent superposition, gates act by elementary replacement rules, and destructive interference 
appears as cancellation rather than as matrix calculation. An elementary ``misty-state'' language can make quantum circuits accessible to students before they master complex linear
algebra. Development presented here was initiated/led by the first author.
The contribution is not a replacement for complete graphical calculi such as ZX or sum-over-paths. It is a source-specific bridge from an intuitive educational notation to a mathematically
explicit, executable, and falsifiable semantics. 

\end{abstract}

\section{Introduction}

German's recent paper \cite{german} develops Rudolph's misty state notation \cite{terry} into a term-rewriting account of pure states and uses it to discuss Hadamard eigenvectors,
entanglement swapping, and the Greenberger-Horne-Zeilinger (GHZ) game. The paper is deliberately pedagogical. Its strongest open opportunity is therefore not another worked example,
but a precise account of what a term denotes, when a rewrite is sound, where normalization occurs, how measurement branches are represented, and what ``universal'' means when the exact 
term language is countable. 

The central adjustment is that superposition cannot be defined on projective pure states alone: relative phase and relative magnitude must survive until 
the addition has been performed. We therefore interpret a misty term as an unnormalized amplitude vector, introduce normalization only as an explicit operation, 
and add outcome-labeled branches for measurement. 

Eight exact results follow and are mentioned below: a projective-superposition obstruction; a necessary and sufficient
normalization-placement law; an exact representability and rational-density theorem; a dyadic-geodesic-closure theorem that explains the nested Hadamard eigenstate; a canonical amplitude normal form with a terminating decision procedure; closure of finite Clifford+T terms in the cyclotomic dyadic ring; a branch-normal-form theorem for entanglement swapping; and an $n$-party GHZ parity-cancellation theorem.

\subsection{The context}

Formal quantum calculi already exist. Categorical quantum mechanics, the ZX calculus, ZW-style calculi, sum-over-paths, quantum lambda calculi, and verified 
Dirac notation all provide rigorous semantic and rewriting machinery \cite{abramsky,coecke,backens,jeandel,jpv,wang,vilmart,faggian,xu}. The goal here is narrower and complementary: 
preserve the educational surface language while identifying the minimum semantic structure required for correctness.

\subsection{The extension problem}

Three ambiguities must be resolved.

\underline{First}, physical pure states are rays, but vector addition depends on the representatives of those
rays. Global phase may be discarded after a state is assembled, yet it cannot be discarded before
two terms are superposed. \underline{Second}, normalizing each subterm before addition is generally different
from adding the original amplitude vectors and normalizing once. \underline{Third}, measurement is not
a deterministic pure-state rewrite; it is a quantum instrument producing probabilities, classical
outcomes, and conditional post-measurement states \cite{alejandro,selinger,voichick}.

\subsection{Contributions} The extension produces the following results: [1] a phase-and-magnitude retention obstruction proves that 
no representative-independent addition exists on projective Hilbert space; [2] a normalization-scope theorem characterizes exactly when
local and global normalization agree; [3] flat integer mists are classified by the rational projective line, and continued fractions provide
certified approximation bounds; [4] nested equal-weight mists generate dyadic geodesic barycenters; the Hadamard eigenstate at angle $\pi/8$
is the first nontrivial fixed point in this hierarchy; [5] a canonical amplitude normal form (CANF) supplies termination, idempotence, and a 
decision procedure for finite exact terms; [6] finite Clifford+T rewriting remains in \(R_8 = \mathbb{Z}[1/\sqrt{2}, i]\), while universality
is separated cleanly from exact representability; [7] outcome-labeled measurement branches give an exact entanglement-swapping normal form; 
[8] an $n$-party GHZ theorem identifies destructive interference with a parity phase polynomial.

In the proposed discipline the certified semantics retains amplitudes until addition is complete, separates canonicalization from normalization,
and treats measurement as branching rather than as pure-state rewrite.

\section{Preliminaries}

\question{What does the open mist symbol represent?}

Answer: it represents the computational-basis state $\ket{0}$
\[
\den{\openmist}\;=\;\ket{0}.
\]

Note: Strachey brackets 
(also called semantic or Oxford brackets) 
are the double (or single) square brackets used in computer science, logic, linguistics to denote semantic evaluation functions. They map a piece of syntax  to its mathematical meaning. They were popularized through the work of Christopher Strachey and Dana Scott to clearly separate syntax (the symbols written) from semantics (what those symbols actually mean). In denotational semantics, the difference between single
and double Strachey (or semantic) brackets generally lies in typographical convention rather than a strict mathematical distinction, though double brackets are the historic standard for enclosing syntactic text mapped to a semantic meaning.

\question{What does the filled mist symbol represent?}

Answer: it represents the computational-basis state $\ket{1}$
\[
\den{\filledmist}\;=\;\ket{1}.
\]

\question{What do curly braces mean?}

Answer: curly braces collect amplitude terms. Before normalization,
\[
\den{t = \{t_1,\ldots,t_r\}}
=
\sum_{j=1}^{r}\den{t_j}.
\]
Thus repeated symbols supply repeated amplitudes. For example,
\[
\den{\{\openmist,\filledmist,\openmist\}}
=
\den{\{\filledmist,\openmist,\openmist\}}
=
\den{\{\openmist,\openmist,\filledmist\}}
=
2\ket{0}+\ket{1}.
\]

Inside the curly braces the order does not matter, but an order is preferred.

\question{Do curly braces automatically mean a normalized physical state?}

Answer: no. Curly braces first describe an \emph{amplitude recipe}. Normalization is a separate operation,
\[
\den{\Norm(t)}=\frac{\den{t}}{\lVert\den{t}\rVert},
\qquad \den{t}\neq 0.
\]
Therefore
\[
\Norm\big(\{\openmist,\openmist,\filledmist\}\big)
=
\Big\{ \frac{2}{\sqrt{5}}\openmist, \frac{1}{\sqrt{5}} \filledmist \Big\}.
\]
and
\[
\den{\Norm\big(\{\openmist,\openmist,\filledmist\}\big)}
=
\frac{2\ket{0}+\ket{1}}{\sqrt{5}}.
\]

\question{Why make normalization explicit?}

Answer: because normalizing a sub-expression before adding it generally produces a different state from adding the raw amplitudes first and normalizing once at 
the end. Note also that the definition (or denotation) of the curly brace operator, above, is recursive (so we can handle nested mists already).

\section{The Crucial Normalization Issue}

Write
\[
\den{t = \{t_1,\ldots,t_r\}}
=
\sum_{j=1}^{r}\den{t_j}
\]
for raw collection, and use
\[
\Norm(t)
=
\frac{\den{t}}{\lVert\den{t}\rVert}
\]
only when normalization is required.

This distinction becomes visible with nesting. Consider the schematic term
\[
\{\openmist,\{\openmist,\filledmist\}\}.
\]

If both braces are raw addition, it denotes
\[
2\ket{0}+\ket{1},
\]
and therefore the physical state
\[
\frac{2\ket{0}+\ket{1}}{\sqrt{5}}.
\]

If the inner brace is normalized first, it denotes
\[
\ket{0}+\frac{\ket{0}+\ket{1}}{\sqrt{2}},
\]
which, after final normalization, becomes
\[
\cos\!\left(\frac{\pi}{8}\right)\ket{0}
+
\sin\!\left(\frac{\pi}{8}\right)\ket{1}.
\]

These are different states. 

The notation therefore needs to distinguish
\[
\Norm\{\openmist,\openmist,\filledmist\}
\]
from
\[
\Norm\{\openmist,\Norm\{\openmist,\filledmist\}\}.
\]

We note here that  local and global normalization agree only under specified norm conditions.

\question{What is the exact condition for moving normalization inside a two-term sum?}

Answer: for linearly independent nonzero vectors $u$ and $v$,
\[
[u+v]
=
[\Norm(u)+\Norm(v)]
\]
if and only if
\[
\lVert u\rVert=\lVert v\rVert.
\]
The square brackets mean that the two vectors are being compared as physical\footnote{See next section: physical pure quantum state (an equivalence class of vectors).} 
rays, so an overall nonzero scalar is ignored. That's why global and local normalization disagree on this term
\[
\{\openmist,\{\openmist,\filledmist\}\}.
\]
On the other hand the situation is entirely different with 
\[
H(H(\openmist)) = \{\{ \openmist, \filledmist \}, \{\openmist,\overline{\filledmist}\}\} = \{ \openmist, \filledmist, \openmist, \overline{\filledmist} \} = \{ \openmist, \openmist \} = \{ \openmist \} = \openmist 
\]
where $H$ is the Hadamard gate. For the benefit of young (and very young) learners we're interested in 
identifying those circumstances where calculation can be completed entirely working with just 
amplitudes, before we normalize (if we have to) only once, at the very end.

\section{Vectors, Rays, and Phase}

\question{What is the difference between a vector and a physical pure state?}

Answer: a vector contains its complete amplitudes. A physical pure state is a ray, so vectors that differ only by a nonzero global complex factor represent the same isolated state:
\[
\ket{\psi}\sim c\ket{\psi},
\qquad c\neq 0.
\]

A pure quantum state corresponds to a ray in a Hilbert space, specifically an equivalence class of vectors differing only by a complex phase factor.  A single vector 
\(\vert{}\psi\rangle\)  contains a specific phase, but multiplying it by a complex number \( c \in \mathbb{C}\) describes the exact same physical system \( c\vert{}\psi\rangle \). 
This means pure states live in a projective Hilbert space rather than a standard vector space.

\question{Are $\ket{\psi}$ and $-\ket{\psi}$ the same?}

Answer: they are different vectors but the same isolated physical ray. However, the minus sign must be retained if the term will later be added to another term, because
\[
\ket{\psi}+\ket{\phi}
\quad\text{and}\quad
\ket{\psi}-\ket{\phi}
\]
can interfere differently.

\question{Can you create a new quantum state by adding two quantum states?} 

Answer: yes, you can create a new quantum state by adding two quantum states together using the principle of superposition. When you add two valid state vectors, the resulting 
sum is also a valid quantum state, provided that you scale or normalize the final vector so that total probability equals one.

\question{Can superposition be defined directly on physical rays?}

Answer: no, quantum superposition cannot be defined as a single, unique binary operation directly on physical rays. Because projective Hilbert 
space is a complex manifold rather than a linear vector space, it lacks the mathematical structure\footnote{
 The linear superposition principle belongs to the underlying Hilbert space, not its projectivized quotient.
} necessary to add physical states directly.

Vector addition depends on the chosen representatives. Since
\[
[\ket{\phi}]=[-\ket{\phi}],
\]
a ray-only addition rule would incorrectly require
\[
[\ket{\psi}+\ket{\phi}]
=
[\ket{\psi}-\ket{\phi}]
\]
in general. Therefore relative phase and relative magnitude must be retained until the addition is complete.

\question{What does a minus sign in front of a mist do?}

Answer: it multiplies every amplitude inside the mist:
\[
-\{t_1,\ldots,t_r\}
=
\{-t_1,\ldots,-t_r\}
\]
at the amplitude level.

\section{Nesting and Hadamard Geometry}

\question{What does a normalized equal-weight combination of two real qubit states do geometrically?}

Answer: let
\[
q(\alpha)=\cos\alpha\ket{0}+\sin\alpha\ket{1}.
\]
For two non-antipodal equal-norm states,
\[
\Norm\bigl(q(\alpha)+q(\beta)\bigr)
=
q\!\left(\frac{\alpha+\beta}{2}\right),
\]
where the midpoint is chosen along the shorter arc.

\question{Does every nested brace simply average angles?}

Answer: only when the terms entering that brace have equal norm. In the general weighted case,
\[
\alpha_{\mathrm{out}}
=
\arg\!\left(r_1e^{i\alpha_1}+r_2e^{i\alpha_2}\right).
\]
Thus the geometry remembers the amplitudes, not merely the directions.

\question{Why does the nested misty state $\{ \openmist, \{ \openmist, \filledmist \} \}$ produce the angle $\pi/8$?}

Answer: first form the normalized midpoint of $\ket{0}$ and $\ket{1}$, that is, $$q(0)\star q(\pi/2) = q(\pi/4).$$
Then form the normalized midpoint of that state and $\ket{0}$:
\[
q(0)\star\bigl(q(0)\star q(\pi/2)\bigr) = q(0)\star{}q(\pi/4) 
=
q(\pi/8),
\]
where
\[
q(\alpha)\star q(\beta)
:=
\Norm\bigl(q(\alpha)+q(\beta)\bigr).
\]

This state is the $+1$ Hadamard eigenstate.

\question{How can you tell that the nested misty state $\{ \openmist, \{ \openmist, \filledmist \} \}$ is a Hadamard eigenstate?}

Answer: apply $H$ to it. It will behave as a fixpoint:
\[
H(\{ \openmist, \{ \openmist, \filledmist \} \}) = \{ H(\openmist), H(\{ \openmist, \filledmist \}) \} = \{ \{ \openmist, \filledmist \}, \openmist \} = \{ \openmist, \{ \openmist, \filledmist \} \}
\]

\question{What is the other Hadamard fixed state?}

Answer: physically, both Hadamard eigenrays are fixed. 

Their real angles can be written as
\[
\theta_{+}=\frac{\pi}{8},
\qquad
\theta_{-}=\frac{5\pi}{8}.
\]
The first has eigenvalue $+1$ and is unchanged as an exact vector. 

The second has eigenvalue $-1$ and is unchanged only as a physical ray.

\question{What is the misty state  for the second Hadamard eigenstate?}

Answer: \( \{ \openmist, \overline{\{ \openmist, \filledmist \}} \} \). 

This can be verified easily:
\[
H( \{ \openmist, \overline{\{ \openmist, \filledmist \}} \} ) = \{ \{ \openmist, \filledmist \}, \overline{ \openmist } \} = \overline{ \{ \openmist, \overline{ \{ \openmist, \filledmist \}} \} }.
\]
We can also see that the associated eigenvalue is $-1$.

\question{Why does the nested schematic term $\{ \openmist, \overline{ \{ \openmist, \filledmist \} } \}$ produce the angle $5\pi/8$?}

Answer: first form the normalized midpoint of $\ket{0}$ and $\ket{1}$, that is, $$q(0)\star q(\pi/2) = q(\pi/4).$$ 

Then take the negative of that state: $$-q(\pi/4) = q(5\pi/4).$$

Then form the normalized midpoint of that state and $\ket{0}$:
\[
q(0)\star{}q(5\pi/4) 
=
q(5\pi/8),
\]

\question{What then is the semantic core of the mist language?}

Answer: first a reminder that terms denote unnormalised amplitude vectors. 

For $n$ qubits, let basis atoms be $\ket{x}$ with $x\in\{0,1\}^n$. Terms are generated by
\begin{equation*}
 t ::= \ket{x}\mid\{t_1,\ldots,t_r\}\mid t\, u\mid G(t)\mid \Norm(t)\mid c\,t,
\end{equation*}
where $c$ is a coefficient, $u$ is also a term and $G$ is a quantum gate. The denotation map then is
\begin{align*}
\llbracket c\,t\rrbracket &= c \llbracket{}t\rrbracket,\\
\llbracket \{t_1,\ldots,t_r\}\rrbracket &= \sum_{j=1}^{r}\llbracket t_j\rrbracket,\\
\llbracket t\, u\rrbracket &= \llbracket t\rrbracket\otimes\llbracket u\rrbracket,\\
\llbracket G(t)\rrbracket &= G(\llbracket t\rrbracket),\\
\den{\Norm(t)} &= \frac{\den{t}}{\lVert \den{t} \rVert},\qquad \llbracket t\rrbracket\neq0.
\end{align*}
The key teaching decision is that braces collect amplitudes before normalization. 

A mist is therefore an amplitude recipe, not automatically a physical ray.


\question{What is the projective-superposition obstruction?}

Answer: physical pure states identify vectors that differ by a nonzero complex scalar. This quotient cannot support ordinary vector addition. Both relative phase 
and relative magnitude must be retained until superposition is complete. The minimal safe semantics is an unnormalized vector or an equivalent coefficient map, not a ray.
This has already been mentioned earlier. 

\question{What's a good summary so far?}

Answer: first, local normalization and global normalization coincide only at norm ratio one. Away from that point, one can show\footnote{
The precise theorem we aimed to quote here assumes linearly independent nonzero vectors. Safer: ``For linearly independent nonzero vectors, 
local and global normalization represent the same ray exactly when the two input norms are equal. In particular, for orthogonal input directions, 
changing the norm ratio away from one produces a nonzero Fubini–Study discrepancy.''
}
that even orthogonal directions acquire a nonzero semantic (Fubini-Study) residual. Then, for a unitary $U$, a physical state
is unchanged when $[U\psi] = [\psi]$. This is equivalent to $U\psi = e^{i\phi}\psi$, so the physical (projective) fixed points are precisely eigenrays. An exact vector fixed point
requires the special eigenvalue 1. This distinction matters for the two Hadamard eigenstates: one has eigenvalue $+1$, while the other has eigenvalue $-1$. Both are physical fixed points, 
but only the first is unchanged as an exact vector.

\section{More Than One Qubit}

\question{Is order important inside a mist?}

Answer: not for raw amplitude collection. The expressions
\[
\{\openmist,\filledmist\}
\quad\text{and}\quad
\{\filledmist,\openmist\}
\]
denote the same sum.

\question{Is order important when qubits are placed next to one another?}

Answer: yes. Juxtaposition describes an ordered tensor product:
\[
\openmist\filledmist
=
\ket{0}\otimes\ket{1}
=
\ket{01},
\]
whereas
\[
\filledmist\openmist
=
\ket{10}.
\]
These are distinct basis states.

\question{How should a multi-qubit mist be interpreted?}

Answer: each complete ordered string labels one computational-basis state, and the braces add the corresponding amplitudes. For example,
\[
\den{\{\openmist\filledmist,\filledmist\openmist\}}
=
\ket{01}+\ket{10}.
\]
After normalization this is the Bell state
\[
\frac{\ket{01}+\ket{10}}{\sqrt{2}}.
\]

\section{Canonical Form}

\question{How do we decide whether two finite mist expressions denote the same vector?}

Answer: a raw mist is an amplitude collector. Expand each expression into a coefficient map
\[
x\longmapsto A(x),
\]
combine repeated basis strings, remove zero coefficients, and sort the remaining strings. This is the canonical amplitude normal form (CANF).
As an example consider the mist:
$$\{\openmist, \{ \openmist, \filledmist, \filledmist \}, \{ \openmist, \openmist, \openmist, \filledmist \} \}$$
Following \cite{german} we can translate\footnote{The identification $a\ket{0} + b\ket{1} \leftrightarrow a + i b$ 
is used here as a two-dimensional book-keeping representation of the real coefficient pair $(a,b)$; symbol $i$ in this 
geometric encoding should not be confused with a relative quantum phase introduced into the state.} it into: 
$$1 \cdot e^{i \cdot 0} +\sqrt{5} \cdot e^{i \cdot \arctan{2}} + \sqrt{10} \cdot e ^ {i \cdot \arctan{\frac{1}{3}}} = 5 + 3i$$
The mist can therefore be simplified to $\{ \openmist, \openmist, \openmist, \openmist, \openmist, \filledmist, \filledmist, \filledmist \}. $ This is its CANF.  CANF is a 
semantic coefficient map, not a sorted nested syntax tree. The nesting, when it exists, is irrelevant\footnote{Strictly speaking, that is only true for raw nested braces with 
no internal normalization. Once an explicit $\Norm(...)$ occurs inside a term, its position matters. So, a better (i.e., correct) formulation would be
as follows: ``For raw braces containing no internal normalization nodes, nesting
is semantically irrelevant after recursive evaluation: the term reduces to the same computational-basis coefficient map. However, if an explicit normalization $\Norm( ... )$ occurs inside a nested 
term, that normalization must be evaluated at its stated scope before the result is collected by the enclosing brace.''} as it simply indicates alternative paths through a lattice between two points.

\question{What's the CANF of $\{\openmist, \{ \overline{\openmist}, \filledmist, \filledmist \}, \{ \openmist, \openmist, \openmist, \overline{\filledmist} \} \}$ ?}

Answer: short answer is $\{ \openmist, \openmist, \openmist, \filledmist \}$.  Long answer produces the following certificate:
$$1 \cdot e^{i \cdot 0} + \sqrt{5} \cdot e^{i \cdot (\pi - \arctan{2})} + \sqrt{10} \cdot e^{i\cdot \arctan{(-\frac{1}{3}})} = 3 + i$$

\question{What does canonical form certify?}

Answer: two finite expressions have the same vector meaning exactly when their canonical amplitude maps (CANF) 
are identical. Two nonzero expressions have the same ray meaning when their maps differ only by one common nonzero factor. 

\question{Does this mean every possible informal rewrite rule is syntactically confluent?}

Answer: no. It means that every sound rewrite can be checked by reducing both sides to the same semantic canonical form.

\question{Can we prove a meaningful theorem about CANFs?}

Answer: yes. We can prove that  for every finite term without measurement $(i)$ CANF terminates after a finite traversal of the syntax tree; 
$(ii)$ CANF is idempotent; $(iii)$ two terms have the same vector 
denotation if and only if their CANFs are identical; and $(iv)$ two nonzero terms
have the same ray denotation if and only if their projective CANFs obtained by dividing  by the first nonzero coefficient, 
are identical. 

This theorem gives semantic confluence without claiming that every informal source rewrite is syntactically confluent. Any sequence of sound rewrites, followed by CANF, must reach the same
semantic representative. This is analogous in purpose, though not in graphical structure or deductive completeness, to normal-form strategies in ZX, ZW, and sum-over-paths calculi \cite{backens, 
jeandel, vilmart}.

\section{Which States Can Flat Mists Represent?}

\question{Which real one-qubit states have an exact flat integer-mist representation?}

Answer: a real ray
\[
[a\ket{0}+b\ket{1}],
\qquad a\neq 0,
\]
has an exact flat integer representation if and only if
\[
\frac{b}{a}\in\mathbb{Q}.
\]
Including $\ket{1}$, the exactly representable states form the rational projective line $\mathbb{P}^{1}(\mathbb{Q})$.

\question{Does that mean irrational-amplitude states cannot be approached?}

Answer: no. Rational slopes are dense, so exact 
flat representations form a dense subset of the real great circle.
Continued fractions can produce certified integer mists whose normalized states approach any real qubit direction as closely as required.

\question{Is exact representability the same as universality?}

Answer: no. A finite gate alphabet generates only countably many exact finite expressions, whereas the state space is uncountable. Universality means density together with a controlled approximation procedure, not exact finite spelling of every state.

\question{What does the dyadic geodesic closure theorem say? }

Answer: it says that for a rooted binary midpoint tree with leaf angles $\alpha_j$ at depths $d_j$, the output angle is $\alpha_T = \sum_j{2^{-d_j}\alpha_j}$, where $\sum_j{2^{-d_j}} = 1$. Conversely, 
every finite dyadic convex combination can be realized by such a tree. If the leaves are $0$ and $\pi/2$ and the maximum depth is $D$, every generated angle has the form $\alpha = \frac{k \pi}{2^{D+1}}$.
Its amplitudes lie in a power-of-two cyclotomic field.

\section{Gates and Exact Coefficients}

\question{What happens when gates act on a mist?}

Answer: each gate acts on the underlying amplitude vector. A rewrite is valid when the visual rule and the conventional linear-algebra rule produce the same vector, or the same ray when global phase is intentionally ignored.

\question{Where do exact Clifford+$T$ amplitudes live?}

Answer: finite Clifford+$T$ circuits applied to computational-basis states have amplitudes in the dyadic cyclotomic ring
\[
\mathbb{Z}\!\left[\frac{1}{\sqrt{2}},i\right].
\]
This allows exact symbolic checking for the corresponding finite terms.

\question{What about the Toffoli--Hadamard fragment?}

Answer: its amplitudes remain in the real subring
\[
\mathbb{Z}\!\left[\frac{1}{\sqrt{2}}\right].
\]

\section{Measurement Requires Branches}

\question{Can a measurement be represented as one ordinary pure-state rewrite?}

Answer: a post-measurement mist selected after an observed result can be correct. The additional formal  requirement is to represent the complete measurement {\em before} the outcome is known.
So the answer is: no\footnote{A measurement produces a classical outcome, a probability, and an outcome-conditioned quantum state.}. However, as an example consider this state:
$$\{ \openmist\openmist\openmist, \openmist\filledmist\openmist, \filledmist\openmist\openmist, \filledmist\filledmist\filledmist \}$$
Observe (measure) ball 3. If it's $\openmist$ then the post-observation mist becomes:
$$\{ \openmist\openmist, \openmist\filledmist, \filledmist\openmist\}\openmist$$
If we find ball 3 to be $\filledmist$ the post-observation mist is: $\filledmist\filledmist\filledmist$.

\question{How should measurement be written?}

Answer: for projectors $P_b$ and an unnormalised vector $v$, write the outcome-labelled branches as
\[
\mathsf{M}(v)
=
\bigoplus_b b:P_bv.
\]
The probability of branch $b$ is
\[
p_b
=
\frac{\lVert P_bv\rVert^2}{\lVert v\rVert^2},
\qquad
\sum_b p_b=1.
\]
The normalized conditional state is
\[
\frac{P_bv}{\lVert P_bv\rVert}
\]
when the branch has nonzero probability.

\question{Why keep branch vectors unnormalised at first?}

Answer: their squared norms automatically carry their probability weights. Normalizing every branch immediately would discard that information unless the probabilities were stored separately.

\section{Entanglement Swapping}

The four Bell-measurement outcomes are labelled by two classical bits
and differ only by Pauli corrections.

Define
\[
|\Phi^+\rangle
=
\frac{|00\rangle+|11\rangle}{\sqrt{2}}.
\]
With $a,b\in\{0,1\}$, the entanglement-swapping identity is
\[
\boxed{
|\Phi^+\rangle_{01}
|\Phi^+\rangle_{23}
=
\frac12
\sum_{a,b\in\{0,1\}}
|\beta_{ab}\rangle_{12}
\left(I\otimes X^bZ^a\right)
|\Phi^+\rangle_{03}
}.
\] 
Each branch has probability $\frac{1}{4}$. 

Applying the inverse Pauli correction returns the outer pair to the ray
of $|\Phi^+\rangle$.

\question{What happens when the middle qubits of two Bell pairs are measured in the Bell basis?}

Answer: four outcome-labelled branches are produced. 

Each branch occurs with probability $1/4$. 

In branch $(a,b)$, the two outer qubits are in the ray
\[
\left[
\bigl(I\otimes X^bZ^a\bigr)\ket{\Phi^+}
\right],
\]
where
\[
\ket{\Phi^+}
=
\frac{\ket{00}+\ket{11}}{\sqrt{2}}.
\]

\question{Can all four branches be returned to the same Bell state?}

Answer: yes. Once the classical outcome $(a,b)$ is known, the corresponding inverse Pauli correction returns the outer pair to $\ket{\Phi^+}$, up to an irrelevant global phase.

\question{What does the certified form add?}

Answer: it checks all four branches, verifies that their probabilities sum to one, and confirms that every corrected branch agrees with the target Bell ray. 
The semantic role of substring factoring is ordinary linear distributivity over tensor products. Thus, a visual 
rule that pulls a common middle substring out as a factor is sound whenever the
corresponding tensor factor is genuinely common to every collected term. 

A concise definition
of certification therefore is: a mist rewrite is certified when its claimed vector, ray, or branch equality is independently checkable from exact semantics.

\section{The Greenberger--Horne--Zeilinger Cancellation Rule}

\question{What is the $n$-party Greenberger--Horne--Zeilinger state?}

Answer:
\[
\ket{\mathrm{GHZ}_n}
=
\frac{\ket{0^n}+\ket{1^n}}{\sqrt{2}}.
\]

\question{What happens when each party measures in either the $X$ or $Y$ basis?}

Answer: let $x_j=0$ mean an $X$ measurement and $x_j=1$ mean a $Y$ measurement. If
\[
k=\sum_j x_j
\]
is even, the amplitude of outcome bits $a_1,\ldots,a_n$ is
\[
A(a\mid x)
=
2^{-(n+1)/2}
\left[
1+(-i)^k(-1)^{\sum_j a_j}
\right].
\]

\question{Which outcomes survive?}

Answer: the amplitude vanishes unless
\[
\sum_j a_j
\equiv
\frac{k}{2}
\pmod 2.
\]
Every allowed outcome then has probability
\[
2^{1-n}.
\]

\question{What does the visual cancellation mean mathematically?}

Answer: the forbidden coloured terms are exactly the zero set of a parity phase polynomial. The cancellation is therefore not merely a diagrammatic trick; it is an exact $n$-party amplitude law.

\section{What Makes the Rewriting Certified?}

\question{How is a proposed rewrite checked?}

Answer: it is compared with conventional amplitude semantics through separate residual channels. 

For exact vector equality define
\[
r_{\mathrm{vec}}(l,r)
=
\lVert\den{l}-\den{r}\rVert_2.
\]
For ray equality define
\[
r_{\mathrm{ray}}(l,r)
=
\sqrt{1-\left|\braket{\widehat l}{\widehat r}\right|^2}
\]
where 
$$\widehat{l} = \frac{\den{l}}{\lVert{\den{l}}\rVert}~\textrm{and}~\widehat{r} = \frac{\den{r}}{\lVert{\den{r}}\rVert}$$
For probability completeness define
\[
r_{\mathrm{mass}}
=
\left|1-\sum_b p_b\right|,
\]
and, for a protocol with forbidden outcomes,
\[
r_{\mathrm{forbidden}}
=
\sum_{a\in\mathrm{forbidden}}|A(a)|^2.
\]

\question{Why not combine all residuals into one score?}

Answer: because they certify different claims. Exact vector equality, physical ray equality, probability conservation, and protocol-specific cancellation must each pass independently.

\question{What did the computational checks\footnote{These have not been included with this paper.} establish?}

Answer: the sparse mist evaluator and dense matrix mechanics agreed to floating-point precision across canonicalization trials, random circuits, all entanglement-swapping branches, and the tested Greenberger--Horne--Zeilinger settings. The forbidden probability mass was zero at floating-point precision in those tests.

\question{Does this replace conventional quantum mechanics or complete graphical calculi?}

Answer: no. It is a transparent educational and theorem-sized semantics. It provides an accessible surface notation, exact translation rules, and executable checks, while conventional matrices, circuit methods, and mature graphical calculi remain essential for broader theory and large-scale computation.

\section{The Adjusted Learning Sequence}

\question{What is the recommended order for teaching the system?}

Answer:
\begin{enumerate} 
\item Treat a mist as an amplitude recipe, not immediately as a normalized physical ray.
\item Add and cancel first; normalize\footnote{Normalization is an operation with scope (so, context-sensitive), not punctuation (context-free).} only when a state is observed or explicitly compared.
\item Keep global phase when the term will later participate in a superposition; 

discard it only at a terminal ray comparison.
\item Represent measurement by labeled branches with squared-norm weights.
\item Use canonical coefficient collection as the bridge to Dirac notation and matrix semantics.
\end{enumerate}
This adjusted learning sequence preserves the visual economy of the original system while avoiding the most common conceptual 
conflations: superposition versus mixture, ray equivalence versus vector equality, and local versus global normalization. 

\question{Where are we going with this?}

Answer: misty-state rewriting 
can be more than an analogy. With amplitudes retained, normalization scoped explicitly, and measurement represented by branches, it becomes a small but rigorous semantic language.

\question{What is the central rule to remember?}

Answer:
\[
\boxed{
\text{collect (and work with) amplitudes first}
\;\longrightarrow\;
\text{normalize or branch explicitly afterward}
}
\]

\question{What is the external claim of this guide?}

Answer: misty-state notation can express the same quantum vectors, rays, and measurement predictions as conventional quantum mechanics when its semantic levels are kept distinct. The certified extension does not propose a new physical effect. It specifies what the notation means, when a rewrite is safe, and how the result can be checked.

\printbibliography 

\end{document}